\documentclass[aps,prb,superscriptaddress,showpacs,citeautoscript,twocolumn]{revtex4}

\usepackage{graphicx}
\usepackage{amsmath}
\usepackage{amssymb}

\newcommand{\kso}{k_{\textrm{so}}}

\newcommand{\phii}{\varphi_{\rm i}}
\newcommand{\bnen}{\begin{equation}}
\newcommand{\eden}{\end{equation}}
\newcommand{\bean}{\begin{eqnarray}}
\newcommand{\eean}{\end{eqnarray}}
\newcommand{\bna}{\begin{array}}
\newcommand{\eda}{\end{array}}
\newcommand{\f}{\frac}

\newcommand{\C}{\mathbb{C}}
\newcommand{\J}{\mathbb{J}}

\renewcommand{\vec}[1]{\text{\boldmath{$ #1 $}}}

\newcommand{\spinor}[2]{\left(\bna{c} #1 \\ #2 \eda\right)}

\begin{document}
\title{Skew scattering due to intrinsic spin-orbit coupling \\
in a two-dimensional electron gas}

\author{A.~P\'alyi}
\affiliation{Department of Physics of Complex Systems, E{\"o}tv{\"o}s
University, H-1117 Budapest, P\'azm\'any P{\'e}ter s{\'e}t\'any 1/A, Hungary}

\author{J.~Cserti}
\affiliation{Department of Physics of Complex Systems, E{\"o}tv{\"o}s
University, H-1117 Budapest, P\'azm\'any P{\'e}ter s{\'e}t\'any 1/A, Hungary}

\date{\today}

\begin{abstract}
We present the generalization of the two-dimensional quantum scattering formalism
to systems with Rashba spin-orbit coupling.
Using symmetry considerations, 
we show that the differential scattering cross section depends on the spin 
state of the incident electron, and skew scattering may arise even 
for central spin-independent scattering potentials.
The skew scattering effect is demonstrated by exact results of a 
simple hard wall impurity model. 
The magnitude of the effect for short-range impurities is estimated using 
the first Born approximation. 
The exact formalism we present can serve as a foundation for 
further theoretical investigations.
\end{abstract}


\maketitle

\section{Introduction}
When an electron scatters on an impurity atom with strong spin-orbit interaction
(SOI) in a solid, the features of the process may depend on the spin state of 
the electron.
The presence of the SOI may result in the asymmetry of the differential 
scattering cross section (commonly referred to as the \emph{skew scattering}),
even though the overall spin-dependent potential of the atom is central
\cite{mott,ballentine}.
In ferromagnetic metals, this particular scattering mechanism and the 
finite equilibrium polarization of the conduction electrons can lead to 
a finite potential drop transverse to an applied electric field, 
even in the absence of magnetic field (anomalous Hall effect).
\cite{smit1,smit2,crepieux}
Similar conditions in non-magnetic materials can induce a spin accumulation 
at the edges of the sample (extrinsic spin Hall effect).
\cite{hirsch,engel-eshe,kato,engel-review}

It is well known that the SOI can also be important in clean semiconductor 
samples lacking inversion symmetry.\cite{winklerbook,dresselhaus,rashbaterm}
For example, a finite electric field perpendicular
to the plane of a two-dimensional electron gas in a quantum well causes
the spin splitting of the conduction subbands.\cite{winklerbook} 
A widely used model Hamiltonian
to describe this intrinsic spin splitting was proposed by 
Rashba:\cite{rashbaterm}
\bnen
\label{eq:rashbahamiltonian}
H_0 = \f{{\bf p}^2}{2m^*} + \f{\alpha}{\hbar}(\sigma_x p_y - \sigma_y p_x),
\eden
valid in the one-band effective-mass approximation.
Here $\sigma_x$ and $\sigma_y$ denote the Pauli matrices, and the parameter
$\alpha$ describes the strength of the spin-orbit coupling.
The second term in \eqref{eq:rashbahamiltonian} is usually referred to as
the Rashba SOI or the Rashba term.

Several studies have been devoted to the physical consequences of the interplay 
of the Rashba SOI and impurity scattering.\cite{engel-review,malshukov,shytov,
engel-polarization,inoue,mishchenko,raimondi}
The formation of a spin-polarized electron cloud around an isolated impurity
has been predicted, resulting from the concurrent presence of the Rashba SOI 
and a constant in-plane electric field.
\cite{malshukov}
The effect of the Rashba term on transport (intrinsic spin Hall
effect) and polarization phenomena in disordered systems has been investigated
using semiclassical\cite{engel-review,shytov,engel-polarization} and 
quantum\cite{inoue,mishchenko,raimondi} methods.
In these studies, the role of impurity scattering in the presence of 
Rashba SOI is treated using various approximate models.
However, it is necessary to go beyond these models if we want to reveal 
the details of the electron scattering characteristic of the presence
of the Rashba coupling.

In this work, we provide the generalization of two-dimensional scattering theory
for the Rashba Hamiltonian $H_0$ via the $S$ matrix formalism.
Using only symmetry considerations, we show that the Rashba term can induce 
skew scattering even if the scattering potential is central and 
spin-independent (e.g. an impurity ion or atom with negligible SOI). 
We demonstrate the skew scattering effect on 
the exactly solvable hard wall impurity model.\cite{walls,yeh}
We prove that the effect appears in the first Born approximation, which is a
major difference compared to the conventional skew scattering mechanism
\cite{ballentine,smit2,engel-review}. 
Based on this result, we propose a modification of the frequently used
isotropic or $s$-wave\cite{swave} approximation, in order to take into 
account the effects caused by the intrinsic spin-orbit coupling. 

\section{General formulation}
Consider the scattering problem governed by the Hamiltonian $H=H_0+V$, where
the only assumption for the scattering potential $V$ is to be zero outside a 
circle of radius $R$.
The cylindrical wave eigenfunctions of the Hamiltonian $H_0$ with energy $E$ are
\cite{rashbabilliard}
\bnen
\label{eq:partwaves}
h_{j\tau}^{(d)}(r,\varphi) =
\sqrt\f{k_\tau}{k}
\spinor{\tau H^{(d)}_{j-\f12}(k_\tau r)e^{-i\varphi/2}}{H^{(d)}_{j+\f12}(k_\tau r)e^{i\varphi/2}}
e^{ij\varphi}, 
\eden
where $k=\sqrt{2m^*E/\hbar^2 + \kso^2}$, $k_\tau = k - \tau \kso$, 
$\kso = \alpha m^*/\hbar^2$, $\tau \in \{\pm 1\}$ is the helicity quantum 
number\cite{yeh},
and $j \in \J$ is the total angular momentum quantum number. 
Here $\J = \{\dots, -\f32, -\f12, \f12, \f32,\dots \}$.
$d \in \{1,2\}$, and
$H^{(1,2)}$ refers to the Hankel function of the first and second kind, respectively
\cite{abramowitz}. 
$h^{(1)}$ ($h^{(2)}$) is an outgoing (incoming) cylindrical wave\cite{yeh}.
The magnitude of the radial particle current carried by the state 
\eqref{eq:partwaves} is independent of $j$, $d$ and $\tau$.
The eigenfunctions of $H_0$ are eigenfunctions of $H$ in the region $r>R$.
Therefore, the elastic scattering of a single incoming partial wave 
$h^{(2)}_{j\tau}$ on the potential $V$ is described by the wave function
\bnen
\label{eq:partwavescat}
\psi_{j\tau} = h^{(2)}_{j\tau} + 
\sum\limits_{j'\tau'} S^{(j'j)}_{\tau'\tau} h^{(1)}_{j'\tau'},
\eden
valid outside the circle of radius $R$.
The sum is for $j'\in\J$ and $\tau' \in \{\pm 1\}$.
The $S$ matrix depends on the actual form of the scattering potential.
Since the partial waves carry the same amount of current, the $S$ matrix is
unitary.
In this section, we derive the relation between the $S$ matrix and
the differential scattering cross section.

The plane wave eigenfunctions of $H_0$ with energy $E$ and helicity 
$\tau \in \{ \pm \}$, propagating in a direction $\phii$ are
\bnen
\label{eq:pw}
\phi_{\tau,\phii}(r,\varphi) 
= \eta_\tau(\phii) e^{ik_\tau r \cos(\varphi-\phii)},
\eden
where 
\bnen
\eta_\tau(\phii) = \f{1}{\sqrt{2}} \spinor{\tau i e^{-i\phii/2}}{e^{i\phii/2}}.
\eden
The partial wave expansion of the plane wave \eqref{eq:pw} is\cite{abramowitz}
\bnen
\phi_{\tau,\phii} = \f{1}{2} \sqrt{\f k {k_\tau}} \sum\limits_{j} i^{j+1/2} 
[h_{j\tau}^{(2)} + h_{j\tau}^{(1)}] e^{-ij\phii}.
\eden
Using this expansion, the principal asymptotic form of the Hankel
functions, and equation \eqref{eq:partwavescat}, it can be shown that the
total wave function describing the scattering of the plane wave 
$\phi_{\tau,\phii}$ asymptotically far from the scatterer is
\bnen
\tilde{\psi}^{\rm (tot)}_{\tau,\phii} = 
\phi_{\tau,\phii} + 
\tilde{\psi}^{\rm (sc)}_{\tau,\phii},
\eden
with
\bnen
\tilde{\psi}^{\rm (sc)}_{\tau,\phii} (r,\varphi) = 
\f{1}{\sqrt{r}}\sum\limits_{\tau'}e^{ik_{\tau'}r}
\eta_{\tau'}(\varphi) f_{\tau'\tau}(\varphi,\phii).
\eden
Here
\bnen
f_{\tau'\tau}(\varphi,\phii) = \f1{\sqrt{2\pi i k_\tau}}
\sum\limits_{j'j} 
e^{ij'(\varphi-\f \pi 2)} F^{(j'j)}_{\tau'\tau} e^{-ij(\phii-\f \pi 2)},
\eden
and 
$F^{(j'j)}_{\tau'\tau} = S^{(j'j)}_{\tau'\tau} - \delta_{j'j} \delta_{\tau'\tau}$.
In the following we will refer to the $2\times 2$ complex matrix $f$ as
the scattering amplitude matrix.

\begin{figure}[t]
\includegraphics[scale=0.4]{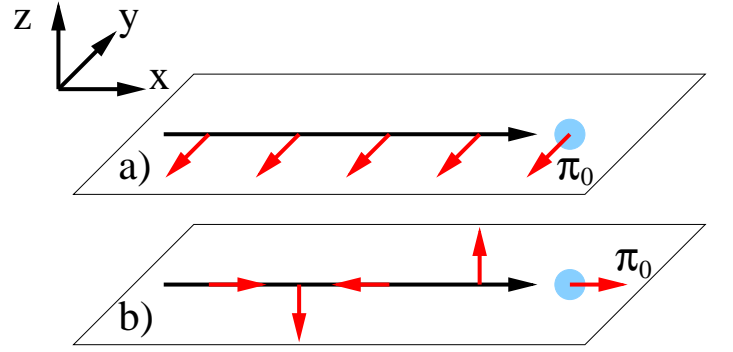}
\caption{\label{fig:spinrotates}
(color online) Demonstration of the spin dynamics of plane waves a) with and 
b) without definite helicity (Eq.~\eqref{eq:superpw}). 
a) $\gamma = (1,0)$, b) $\gamma = (1,i)/\sqrt2$.
In both figures, the black arrow denotes the direction of propagation ($\phii = 0$), 
the red arrows denote the local polarization vector, and the circle denotes the origin.
$\vec{\pi}_0$ is the polarization vector in the origin.}
\end{figure}

A plane wave with a given energy and propagation direction is not necessarily
a helicity eigenstate. 
Coherent superpositions of the two helicity-eigenstate plane waves are described
by the wave function
\bnen
\label{eq:superpw}
\phi_{\gamma,\phii} = \sum\limits_{\tau} \phi_{\tau,\phii} \gamma_\tau,
\eden
where $\gamma = (\gamma_+,\gamma_-)^{\rm T} \in S_1(\C^2)$ 
(the unit circle in the usual $\C^2$ Hilbert space).
The spin dynamics of such a superposed plane wave is essentially the same as 
in a Datta-Das spin transistor\cite{dattadas}.
We show the spin dynamics of two examples in Fig.~\ref{fig:spinrotates}.

In order to consider the scattering of the plane waves without definite 
helicity in \eqref{eq:superpw}, 
we introduce the $2\times2$ complex matrix
\bnen
D(r,\varphi) = 
\left(e^{ik_+ r}\eta_+(\varphi),e^{ik_- r} \eta_-(\varphi)\right).
\eden
Note that for every $r$ and $\varphi$, $D$ is unitary.
With this notation, the scattered part of the wave function describing the
scattering process of the superposed plane wave \eqref{eq:superpw} is
\bnen
\label{eq:scatwave}
\tilde{\psi}^{\rm (sc)}_{\gamma,\phii}(r,\varphi) = 
\sum\limits_{\tau} \tilde{\psi}^{\rm (sc)}_{\tau,\phii}(r,\varphi) \gamma_{\tau} =
\f{1}{\sqrt{r}}D(r,\varphi) f(\varphi,\phii) \gamma
\eden

The differential scattering cross section is
\bnen
\sigma_{\rm diff}(\varphi,\phii;\gamma) = 
r | \tilde{\psi}^{\rm (sc)}_{\gamma,\phii} (r,\varphi) |^2.
\eden
Since $D$ is unitary, the differential cross section does not depend on $r$, as 
it is expected:
\bnen
\sigma_{\rm diff}(\varphi,\phii;\gamma) =
\gamma^\dag [f(\varphi,\phii)]^\dag f(\varphi,\phii) \gamma.
\eden
In order to get a more transparent formula, we take the expansion of
the scattering amplitude matrix using the unit matrix $\sigma_0$ and the 
Pauli matrices $\sigma_1$, $\sigma_2$, $\sigma_3$:
\bnen
f(\varphi,\phii) = \sum\limits_{m=0}^{3} u_m(\varphi,\phii) \sigma_m.
\eden
The differential scattering cross section in terms of these coefficients $u$
is
\bnen
\label{eq:dcs}
\sigma_{\rm diff}(\varphi,\phii;\gamma) = 
c(\varphi,\phii) + 
{\bf v}(\varphi,\phii) \cdot {\bf P}(\gamma),
\eden
with 
\bean
\label{eq:cdef}c &=& \sum\limits_{m=0}^{3} |u_m|^2, \\ 
\label{eq:vdef}{\bf v} &=& 2 \ {\rm Re} (u_0^* {\bf u}) - i({\bf u} \times {\bf u}^*), \\
\label{eq:pdef}
{\bf P}(\gamma) &=& \gamma^\dag \mbox{\boldmath{$\sigma$}} \gamma.
\eean
Here $\vec{\sigma} = (\sigma_x,\sigma_y,\sigma_z)$, therefore ${\bf P}(\gamma)$
is a three-dimensional unit vector for arbitrary $\gamma\in S_1(\C^2)$.
In equation \eqref{eq:dcs} the differential scattering cross section corresponding
to the angle $\varphi$ is written in terms of the $S$-matrix 
(hidden in $c$ and ${\bf v}$), the propagation direction of the incoming plane 
wave $\phii$ and the complex vector $\gamma$ characterizing the superposed,
non-helicity-eigenstate plane wave. The result \eqref{eq:dcs} is completely
general, valid for arbitrary finite-range scattering potentials. 
This formula will play a central role in the derivation of the main
result of our paper.

To substitute the rather formal quantity $\gamma$ with a real physical quantity
in \eqref{eq:dcs}, we can use the polarization vector of the superposed plane 
wave in the origin,
\bnen
\vec{\pi}_0 (\gamma,\phii) = {\bf P}\left(\phi_{\gamma,\phii}({\bf r} =0)\right)
\eden
instead.
Note that the ${\bf P}$ function is defined in \eqref{eq:pdef}.
It can be shown that the following relation holds between the formal quantity 
${\bf P}(\gamma)$ and the real physical quantity $\vec{\pi}_0(\gamma,\phii)$ :
\bnen
\label{eq:rotation}
\vec{\pi}_0(\gamma,\phii) = \left(\begin{array}{ccc}
0 & \cos\phii & \sin\phii \\
0 & \sin\phii & -\cos\phii \\
-1 & 0 & 0
\end{array}\right) \cdot {\bf P}(\gamma).
\eden
This relation can easily be checked for the examples shown in 
Fig.~\ref{fig:spinrotates}.
Equation \eqref{eq:dcs} together with \eqref{eq:rotation} expresses the 
differential scattering cross section corresponding to the angle $\varphi$ 
as the function of the plane wave propagation direction $\phii$, the $S$-matrix 
and the polarization vector of the incoming plane wave in the origin $\vec{\pi}_0$.

\section{Simple scattering potentials}
The derivation so far has been completely general. Now we restrict our analysis
to special scattering potentials. We call the $V$ scattering potential simple, 
if it preserves the three fundamental symmetries of the Rashba Hamiltonian
$H_0$: time reversal ($i\sigma_y C$, where $C$ is the complex conjugation),
rotation around the $z$ axis ($J_z=-i\hbar\partial_\varphi + \hbar \sigma_z/2$) 
and the combined symmetry of real space reflection and
spin rotation ($\sigma_y P_x$, where $P_x$ is the spatial reflection with respect 
to the $x$ axis).
It is clear that the spin-independent central potentials are simple.

For simple scattering potentials, the three symmetry operations above 
are compatible with $H$, which will result in a remarkable simplification
of the $S$ matrix and, therefore, of $\sigma_{\rm diff}$.
It can be shown that the consequence of the time reversal, rotational and 
combined symmetries, respectively:
\bean
S^{(j'j)}_{\tau'\tau} &=& \tau'\tau e^{i(j-j')\pi} S^{(-j,-j')}_{\tau\tau'},\\
S^{(j'j)}_{\tau'\tau} &=& \delta_{j'j} S^{(jj)}_{\tau'\tau},\\
S^{(j'j)}_{\tau'\tau} &=& \tau'\tau e^{i(j-j')\pi} S^{(-j',-j)}_{\tau'\tau}.
\eean
As a consequence of these relations, the scattering amplitude matrix --
and hence every quantity derived from that -- depends only on
the scattering angle $\theta = \varphi-\phii$.
The explicit form of $f$ for simple scattering potentials:  
\bean
f_{\tau,\tau}(\theta) &=& 
\sqrt{\f{2}{\pi i k_\tau}} \sum\limits_{j \in \J^+} \cos (j\theta) F^{(jj)}_{\tau,\tau},\\
f_{-\tau,\tau}(\theta) &=& 
\sqrt{\f{2 i}{\pi k_\tau}} \sum\limits_{j \in \J^+} \sin (j\theta) F^{(jj)}_{-\tau,\tau}.
\eean
Note that the diagonal (off-diagonal) elements of the scattering amplitude 
matrix $f$ are even (odd) functions of the scattering angle $\theta$.

By definition, skew scattering is absent in the process if the differential 
cross section $\sigma_{\rm diff}$ in \eqref{eq:dcs} is an even function of 
the scattering angle $\theta$ for every $\gamma \in S_1(\C^2)$.
Using the symmetry properties of the components of the scattering amplitude 
matrix $f$, one can show that $c$ and $v_3$ are even functions of 
$\theta$. On the other hand, $v_1$ and $v_2$ are odd. 
It means that the absence of skew scattering is not provided by the symmetry
properties of the total Hamiltonian even if the scattering potential is simple. 
This is the main result of our paper.

To be a bit more specific, we can say that if the incoming plane wave has 
a definite helicity, i.e. $\gamma \propto (1,0)$ or $\gamma \propto (0,1)$, then
using \eqref{eq:pdef} we get ${\bf P}(\gamma) \parallel (0,0,1)$, therefore 
equation \eqref{eq:dcs} and the even character of $c$ and $v_3$ implies the 
absence of skew scattering.
On the other hand, if the incoming plane wave is a finite superposition of the
two helicity eigenstates (i.e.~both components of $\gamma$ are finite),
then the skew scattering effect arises if $v_1$ or $v_2$ are finite.
We note that obviously our symmetry considerations are not capable to tell
whether $v_1$ or $v_2$ is finite or not -- having a specific $V$ potential in
hand, we have to solve the scattering problem and calculate the elements
of the $S$-matrix in order to learn the answer.

\section{Hard wall impurity model}
In order to demonstrate the predicted skew scattering effect, we present 
exact results for a hard wall potential:
\bnen
V(r) = \left\{\bna{cl}0 & \mbox{if }r>R \\ \infty & \mbox{otherwise}\eda\right.
\eden
This potential $V$ is simple. 
We refer to Refs.\, \onlinecite{yeh} and\,\onlinecite{walls} for the derivation 
of the elements of the $S$ matrix.

\begin{figure}
\rotatebox{-90}{
\includegraphics[scale=0.72]{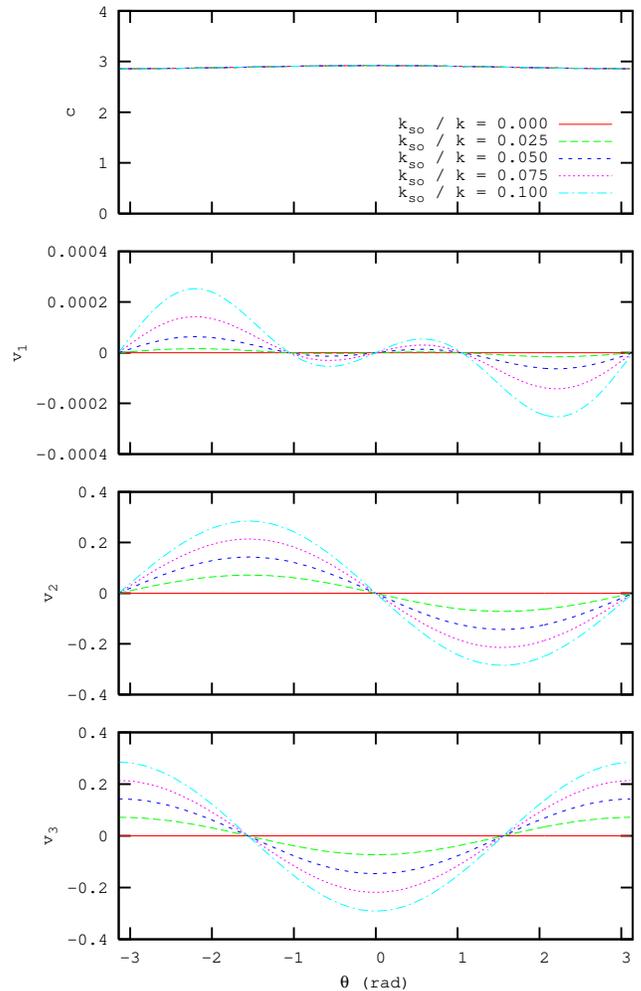}}
\caption{(color online) The quantities $c$, $v_1$, $v_2$ and $v_3$ (in units of 
$R$) determining the differential scattering cross section 
(see eq. \eqref{eq:dcs}) as functions of the scattering angle 
$\theta =\varphi-\phii$ for a central hard wall potential. $kR$ = 0.04. 
\label{fig:candv}}
\end{figure}

We focus on the low-energy properties of the scattering, i.e. $kR \ll 1$, 
because this limit corresponds to the most widely studied short-range impurity 
models\cite{inoue,mishchenko,raimondi}. 
If we consider the scattering of electrons at the Fermi energy, then it
is realistic to set the parameter $\kso/k$ between zero and $0.1$\cite{grundler}.
Exact results for such parameter values are shown in Fig.~\ref{fig:candv}, 
where we have plotted the angular dependence of the key quantities 
$c$, $v_1$, $v_2$ and $v_3$ defined in eqs. \eqref{eq:cdef} and \eqref{eq:vdef}.
Apparently, $v_1$ and $v_2$ does not vanish for finite $\kso$, therefore
skew scattering is present in the process indeed.
Other features of the results are summarized as:
(i) the quantity $c$ is practically independent of $\kso$; 
(ii) for $\kso=0$ we have ${\bf v} =0$, therefore the differential cross section 
(which is equal to $c$ in this case) is spin-independent, symmetric in $\theta$
and approximately a constant function of $\theta$;
(iii) for finite $\kso$, the magnitude of $v_1$ is much smaller than that of
$v_2$ and $v_3$; the latter ones appear to have the same magnitude for a 
given $\kso$, and their magnitude seems to scale linearly with $\kso$.

\begin{figure}
\rotatebox{-90}{
\includegraphics[scale=0.3]{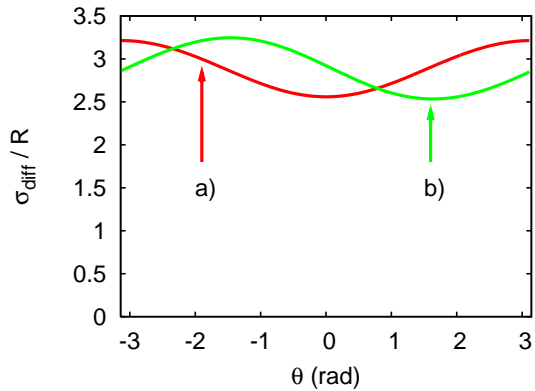}}
\caption{(color online) Differential scattering cross sections exactly calculated
in the hard wall impurity model, corresponding to the incoming plane waves of 
Fig.~\ref{fig:spinrotates}: a) $\gamma=(1,0)$, b) $\gamma = (1,i)/\sqrt 2$.
$k R = 0.04$ and $k_{\rm so}/k = 0.1$.
\label{fig:crosssection}}
\end{figure}

We present the exact differential cross sections calculated using \eqref{eq:dcs}
in Fig.~\ref{fig:crosssection}, corresponding to the two example plane waves of 
Fig.~\ref{fig:spinrotates}.
The $\sigma_{\rm diff}$ of the plane wave with definite helicity (a) is 
symmetric, but for the superposed plane wave (b) the skew scattering effect
is clearly visible even for the realistic value of $k_{\rm so}/k = 0.1$.

In order to understand the features (i) -- (iii) 
of the exact results for the key quantities 
$c$ and ${\bf v}$ (shown in Fig.~\ref{fig:candv}), we calculate
them in the first Born approximation for the simplest 
potential modeling short-range impurites: $V({\bf r}) = \kappa \delta({\bf r})$,
where $\delta$ is the Dirac-delta and $\kappa$ represents the strength of the
potential.
For the incoming plane wave in \eqref{eq:superpw}, the scattered wave
within this approach is\cite{mott,ballentine} 
\bnen
\psi^{\rm (sc)}_{\gamma,\phii} = G^+_E V \phi_{\gamma,\phii},
\eden
where $G^+_E$ is the retarded Green's function of the Rashba Hamiltonian $H_0$.
The exact form of the Green's function in position representation is known.
\cite{walls,rashbabilliard}
Exploiting the simple form of our potential, we find
\bnen
\psi^{\rm (sc)}_{\gamma,\phii}({\bf r}) = 
\kappa G^+_E({\bf r},0) \phi_{\gamma,\phii}(0),
\eden
where $G^+_E({\bf r},{\bf r}')$ is the position matrix element of $G^+_E$.
The first step to derive the scattering amplitude matrix is
taking the $|{\bf r}| \to \infty$ limit of the actual form of 
$G^+_E({\bf r},0)$, and calculating $\tilde{\psi}^{\rm (sc)}_{\gamma,\phii}$.
After that, one can derive the components of $f$ using \eqref{eq:scatwave}.
With some algebra one gets the following results with respect to $c$ and 
$\bf v$:
\bean
\label{eq:cres}
c(\theta) &=& \f{(m^* \kappa)^2}{2\pi \hbar^4 k} = c_0,\\
v_1(\theta) &=& 0,\\
v_2(\theta) &=& -\sin(\theta) c_0 \f{\kso}{k},\\
\label{eq:vres} v_3(\theta) &=& -\cos(\theta) c_0 \f{\kso}{k}
\eean
The qualitative similarity between these results and the exact ones in 
Fig \ref{fig:candv} is remarkable. 
Apparently, the results of this simple short-range impurity model grasp 
all the features (i), (ii) and (iii) of the exact results for $kR\ll 1$ 
listed 
before.
For $\kso=0$ we recover the isotropic, spin-independent differential scattering
cross section $\sigma_{\rm diff}(\theta)=c_0$, as it is expected. 
Including only one further parameter $\kso/k$, our results 
(\ref{eq:cres}-\ref{eq:vres}) provide a generalization of the one-parameter 
isotropic model for scattering in the presence of Rashba SOI.

\section{Summary}
The knowledge of the properties of electron scattering can be the starting point 
for further theoretical predictions of impurity-related solid state phenomena.
The theory of effects related to single, isolated impurities, like the Friedel oscillation, 
Landauer's charge dipole\cite{dittrich} and the spin cloud predicted by 
Mal'shukov and Chu\cite{malshukov} can be based on the knowledge of the 
scattering amplitude matrix $f$. 
In Boltzmann transport theory, the transition probabilities are needed as an
input information in the evaluation of the collision integral\cite{ashcroft}.
Hence the exact formalism presented in this paper can serve as a firm foundation 
of further theoretical investigations.
The two-parameter model for short-range impurities provides a convenient tool
to replace the isotropic approximation in systems with significant Rashba SOI.

In conclusion, we have generalized the formalism of two-dimensional elastic
quantum scattering to systems with finite Rashba spin-orbit coupling. 
Based on symmetry considerations, we have shown that the differential
scattering cross section becomes spin-dependent and can show the skew 
scattering effect even if the scattering potential is central and 
spin-independent.
We have demonstrated the skew scattering by exact results of the 
hard wall impurity model. 
We derived the differential cross section in the first Born approximation for 
a Dirac delta scattering potential, and found remarkable similarity between 
these approximation and the exact results for low scattering energies.
Using the simple formulas gained from the Born approximation, we proposed
a two parameter model to substitute the isotropic or $s$-wave model
of short-range impurity scattering in the presence of Rashba coupling.

We acknowledge fruitful discussions with Cs.~P\'eterfalvi.
This work is supported by European Commission Contract No.~MRTN-CT-2003-504574.


\begin{thebibliography}{xx}
\bibitem{swave}
Strictly speaking, the term `$s$-wave scattering' is paradoxical in the context
of the Rashba Hamiltonian: $s$ would refer to 
partial waves with zero orbital angular momentum, 
however, orbital angular momentum is not a good quantum number if the Rashba 
strength $\alpha$ is finite.
\end{thebibliography}
\end{document}